\documentclass[conference]{IEEEtran}

\usepackage{cite}      

\usepackage{graphicx}  

\usepackage{subfigure} 

\usepackage{url}       

\usepackage{amsmath}   
\newcommand{\eh}[1]{\,\mathrm{#1}}

\newcommand{\ttt}[1]{\cdot10^{#1}}

\newcommand{\dg}{^{\circ}}

\newcommand{\prc}{\eh{\%}}

\newcommand{\lgt}{\log_{10}}

\newcommand{\mr}[1]{\mathrm{#1}}

\renewcommand{\epsilon}{\varepsilon}

\begin{document}

\title{First Results from the IceTop Air Shower Array}

\author{\authorblockN{Stefan Klepser\authorrefmark{1} for the IceCube
Collaboration\authorrefmark{2}}
\authorblockA{\authorrefmark{1}DESY, D-15735 Zeuthen, Germany}
\authorblockA{now at: IFAE Edifici Cn., Campus UAB, E-08193 Bellaterra, Spain, Email: klepser@ifae.es}
\authorblockA{\authorrefmark{2}see www.icecube.wisc.edu}
}


%


\maketitle

\begin{abstract}
IceTop is a 1\,km$^2$ air shower detector presently under construction as a
part of the IceCube Observatory at South Pole. It will consist of
80 detector stations, each equipped with two ice Cherenkov tanks, which
cover 1\,km$^2$. In 2008, the detector is half completed.
One of the design goals of the detector is to investigate cosmic
rays in the energy range
from the knee up to approaching 1\,EeV and study the mass composition of primary
cosmic rays.

In this report the performance of IceTop, the shower
reconstruction algorithms and first results, obtained with one month of data
with an array of 26 stations
operated in 2007, will be presented. Preliminary results are shown for
the cosmic ray energy spectrum in the range of 1 to 80\,PeV. Being located 
at an atmospheric depth of only 700\,g/cm$^2$ at the South
Pole, a high sensitivity of the zenith angle distribution to the mass
composition is observed.

The main advantage of IceTop, compared to other detectors in this energy
range, is the possibility to measure highly energetic muons from air showers in
coincidence with the IceCube detector. The muon rate at a given air shower
energy is sensitive to mass composition. The prospects of this method and
alternative me\-thods to scrutinise different composition models will be
presented.
\end{abstract}


%
\IEEEpeerreviewmaketitle

\section{Introduction}

Cosmic rays in the PeV to EeV energy regime, where the transition from galactic
to extragalactic cosmic rays is expected, are studied by detecting extensive air showers
(EAS) they produce in the atmosphere. In its maximum in terms of particle
number, an EAS predominantly consists of electromagnetic particles. 
IceTop~\cite{icetop_icrc07}, located at $700\eh{g/cm^2}$ on the
south polar glacier, is built to detect showers from
cosmic rays in that energy regime close to their maximum. It is
built on top of the IceCube detector~\cite{icecube_firstyear, icecube_icrc07}, which is located
between $1450$ and $2450\eh{m}$ depth. IceCube is able to detect the light from the
bundles of highly energetic muons in the cores of the EAS. The sizes of
electromagnetic and muonic components of EAS can be used to draw conclusions
on the composition of the primary particles and/or the particle physics that
takes place in the beginning of the shower development. The main difference of
IceTop/IceCube compared to other, mostly surface-bound, EAS arrays is the sensitivity of deep
IceCube to early interaction processes, and the fact that the IceTop signal 
on the surface is predominantly created by electromagnetic shower particles.
This complementary setup may therefore verify existing measurements or cancel out systematic
discrepancies between them, which are for instance caused by the hadronic interaction models used in the
simulation of EAS events.

Furthermore, studies can be done with IceTop alone,
using different inclinations to study composition and the energy
spectrum. Also, efforts are being put into the identification of
single muons at high distances from the shower core, both in IceCube and
IceTop. This may also allow for conclusions on the interaction models or composition.

Another
physics goal not discussed in the following is the use of IceTop in the
context of heliospheric physics~\cite{icetop_helio}.

\section{The IceTop Detector}

\begin{figure}
\centering
\includegraphics[width=0.99\columnwidth]{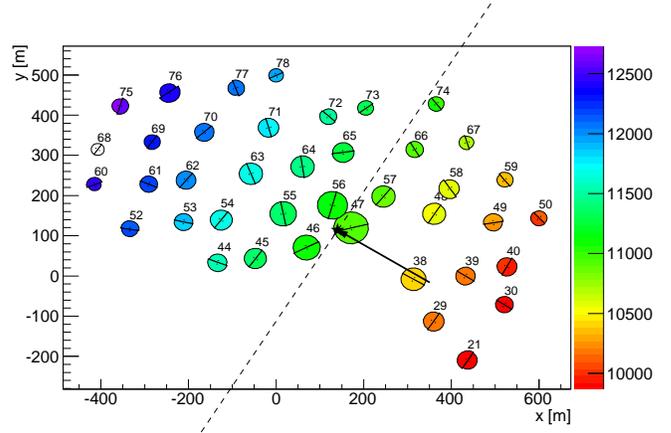}
\caption{Display of a shower event with an energy of about $100\eh{PeV}$ as
recorded with the 40 stations of the 2008
detector. The colors of the halfcircles indicate the pulse times in the tanks,
the sizes scale with the integrated charges. The arrow and the orthogonal
dashed line display the reconstructed direction.}
\label{event}
\end{figure}

In 2007, when the data presented in this paper were taken, IceTop consisted
of 26 detector stations on a triangular grid with a
mean distance of $125\eh{m}$. Each station comprises two $1.86\eh{m}$ diameter
tanks filled with ice to a depth of $90\eh{cm}$.
In each tank, two digital optical modules (DOMs)
detect Cherenkov photons emitted by charged particles in air showers. The DOMs
are mounted on top of the ice bulk, with their light sensitive
halves frozen to the ice surface. A DOM is a light detection
unit that contains a $10''$ photomultiplier tube
(PMT) and electronics to digitise recorded pulses with a precision of
$3.3\eh{ns}$ for about $422\eh{ns}$.
Figure \ref{event} shows the display of an event recorded with the
2008 detector configuration.

The light in the tanks can be reflected multiple times by the inside layers of
the tank walls and
may be recorded by
one or both DOMs, depending on the pulse height and the DOM and trigger
configuration. In 2007, the DOMs were run with two different gains to
enhance the dynamic range. This lead to effective thresholds of about $20$
and $200\eh{PE}$, respectively. In 2008, the gain difference was slightly increased.

\subsection{Trigger and Calibration}

To initiate the readout of DOMs, a coincidence of the two high gain DOMs
of a station is required. Low gain DOMs are read out only if this local
coincidence is already established. The data is written, and thus available
for analysis, if the readouts of six DOMs are launched by a local coincidence. In
2007, the event rate with this configuration was about $14\eh{Hz}$.

The low level processing and calibration
of the data is done in several steps. First, an arrival time is
defined by the leading edge of the pulse and the integrated charge of the
pulse is converted into a number of
equivalent photo electrons (PE). Making use of the muon calibration method~\cite{vemcallevent},
these values are further converted to vertical equivalent muons
(VEM), which makes the analysis essentially independent of
the exact simulation and understanding of the tank and ice properties, which
otherwise would lead to high uncertainties.

The recording of waveforms in principle allows for sophisticated
analyses, exploiting the information in the time structure of the pulse shapes
to investigate the shower structure or particle content. At present, this is not being used.

\section{Shower Reconstruction}

The data sets recorded with IceTop comprise a set of arrival times and
calibrated signal
sizes in units of VEM.
Likelihood maximisation methods are used to reconstruct the location,
direction and size of the recorded showers. In general, the arrival times
contain the direction information and the charge distribution is connected to
size and location of the shower centre. In practice, it turned out to
be a stable and capable approach to start from simple first guess
estimations of direction and shower core and iterate further with detailed
likelihood functions. This also allows an eventual sensitivity of the arrival
times to work on the core location.

\subsection{Fit Procedure and Data Cuts}

The iterative process starts off with the analytic
direction calculation under the assumption of a plane shower front, and the
centre of gravity of the square root of charges (COGSC) as a seed for the
shower core. Then a fit to the lateral distribution of charges
is performed, keeping the
direction fixed. If the core is found closer than $11\eh{m}$ to a
station, the pulses of that stations are discarded and the fit is repeated.
In the next step, a combined fit, using times, charges and a
more realistic curved shower front assumption, leads to the final direction
estimation. In this step, for stability reasons, the direction
is kept flexible only in a limited range. Finally, the lateral function is
fitted again with fixed direction to yield the shower size, energy and lateral
power index results.

In this analysis, we require 5 or more triggered stations to
ensure small errors on the fitted quantities. This
leads to an effective reconstruction threshold (assuming a step function
acceptance) of about $500\eh{TeV}$. A
constant efficiency is reached at about $1\eh{PeV}$, depending on inclination.

The presently applied data cuts mainly assure the convergence of the fits and
the containment of the events inside the array borders. The latter is achieved
not only by requiring the fitted core position to be $50\eh{m}$ (about half a
station distance) inside the array, but in addition asking the COGSC
and the station with the highest charge to fulfill the same condition.
The effective area of the 2007 array, reached with these cuts, is between
$0.094$ and $0.079\eh{km^2}$ for zenith angles between $0\dg$ and $46\dg$ in
the energy range of constant acceptance.

\subsection{Direction and Core Position}

The final event direction is determined under the assumption of a fixed time
delay profile relative to a plane shower front:
\begin{equation}\label{eq:likelihood_time_details}
  \begin{split}
    \Delta t(r_i) &= 19.41\eh{ns}\ [e^{-\left(\frac{r_i}{118.1\eh{m}}\right)^2}-1]
     - 4.823\cdot10^{-4}\eh{\frac{ns}{m^{2}}}\ r_i^2\\
    \sigma_t(r_i) &= 2.92\eh{ns} + 3.77\cdot10^{-4}\ r_i^2.
  \end{split}
\end{equation}
Here, $\Delta t(r_i)$ is the expectation value of the time delay at a
perpendicular distance from the shower axis  $r_i$, and
$\sigma_t(r_i)$ is the expected (Gaussian) standard deviation at that radius. 
This shape was determined by fitting deviations from the fitted plane in
experimental data. The radii $r_i$ depend on the core and direction parameters, so
the fit is in general sensitive to both.
The $68\prc$ resolution that is achieved is $1.5\dg$
and almost independent of energy and zenith angle.

The core position is determined after a lateral fit using the function
introduced in~\cite{ldf_icrc}:
\begin{equation}\label{eq:dlp}
          S(r) =
S_{\mathrm{ref}}\left(\frac{r}{R_{\mathrm{ref}}}\right)^{-\beta_{\mathrm{ref}} - \kappa\,
                   \log_{10}\left(\frac{r}{R_{\mathrm{ref}}}\right)}
\end{equation}
where $r$ again
is the perpendicular distance to the shower axis, $S_{\mathrm{ref}}$
the signal expectation at a distance
$R_{\mathrm{ref}}$, $\beta_{\mathrm{ref}}$ a slope parameter related to the shower age, and
$\kappa$ a (lateral) curvature. In the fit, $R_{\mathrm{ref}}=R_{\mr{grid}}=125\eh{m}$ is
used, leading to a shower size $S_{125}$ and a power index $\beta_{125}$ at that
radius. $\kappa=0.303$ was found constant in simulations and remains fixed in
the fit.

The performance of the fit, in terms of likelihood
distributions and retrieved parameter confidence intervals, is well in agreement with
simulation. The achieved core resolution improves with energy,
approaching $9\eh{m}$ at $3\eh{PeV}$ for zenith angles below $30\dg$.

\begin{figure}
\centering
\includegraphics[width=0.99\columnwidth]{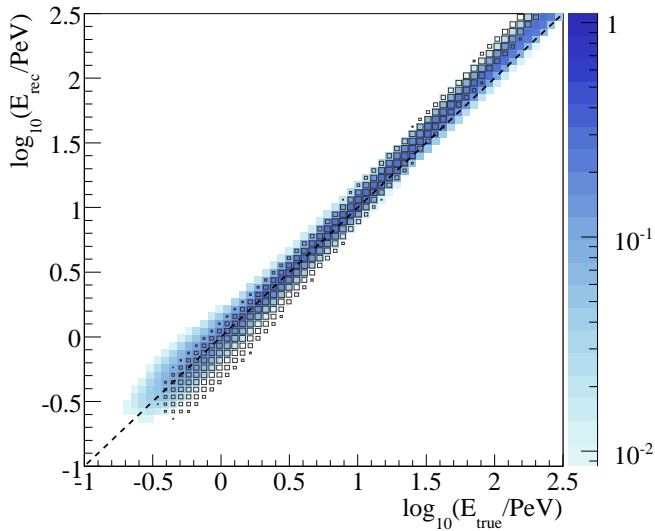}
\caption{Graphical display of the proton (blue) and iron (black) response matrices for air
showers below $30\dg$ zenith angle.}
\label{response_matrix}
\end{figure}

\subsection{Energy Reconstruction}

A simplified simulation study with proton showers was done to derive a
functional form of an energy estimator
$E(S_{\mr{ref}}, R_{\mr{ref}})$ for any given combination
of shower size $S_{\mr{ref}}$ and reference
radius $R_{\mr{ref}}$ (see also \cite{ldf_icrc}).
This was done because it allows us to chose the radius at which the shower size
$S_{\mr{ref}}$ is defined, and the energy is
extracted, for each event individually. At the radius where $S_{\mr{ref}}$ is
independent from the power index
parameter $\beta$, the uncertainty on $S_{\mr{ref}}$, and on the
extracted energy estimator, is minimal.
In the ideal case of a power law, this optimal radius is
the mean of logarithmic radii of all fitted
data points, $\overline{\log{r}}$. 
%
Consequently, to minimise the (statistical) error on the
energy, $S_{\overline{\log r}}$ is calculated for each
event, using Eq.\,\ref{eq:dlp}, and the energy estimator is derived from that.

%

The energy resolution improves with energy and approaches $0.05$ in $\lgt E$,
or $12\prc$ in $E$, at $\sim 3\eh{PeV}$ for zenith angles below $30\dg$. A
graphical display of the resulting response matrix can be seen in
Fig.\,\ref{response_matrix} for proton and iron nuclei.
The faster development of showers from heavy primaries leads to a tilt of the
bands against the diagonal of the matrix.
Since IceTop is close to the shower maximum, the center of rotation,
i.e. where the two bands cross,
lies within the observed energy range. This means that at low energies,
showers from heavy primaries look less energetic
than proton showers, whereas at high
energies they appear more energetic. It shall be noted that the point of
rotation depends on many factors, such as the chosen energy extraction radius
and inclination.

The deviation of the proton response from the diagonal at high energies is connected to
inaccuracies of the simplified simulations with respect to the full detector
simulation. It is corrected by the unfolding.

\section{Studies with IceTop Alone}

IceTop can be used as a standalone air
shower detector, which allows for an early verification of the above
techniques, analysing showers with zenith angles up to $46\dg$.
The different attenuation of proton and iron showers, and its dependence on
the zenith angle, leads to a deviation from the expected isotropic flux if
an incorrect primary composition is assumed.
In this way, IceTop alone is sensitive to composition~\cite{klepser_thesis}. 

%

\subsection{Unfolding Techniques}

The response matrix is defined in a way to relate the true energy spectrum 
to the measured distribution of
first guess energies. It depends on the primary type and
zenith angle. In the case of IceTop, the matrix is only two-dimensional, close
to diagonal and the resolution does not vary much with energy. The unfolding of the spectrum,
which essentially corrects for resolution and an eventual shift, is therefore
not too difficult
and was done with two iterative methods. One is a Bayesian approach
as presented in~\cite{dagostini_unfolding}, the other one is the Gold
algorithm~\cite{gold_unfolding}. To determine correct error bands, a
bootstrap method was used~\cite{bootstrap}, which randomises
the measured distributions
within their error bands, analysing the resulting variations in the
unfolded spectrum. The iteration depths were
adjusted in simulation in a way that the deviation between unfolded spectrum
and assumed true spectrum was minimised in the energy range of interest.

Both algorithms and the error determination were verified in simulation. The
uncertainties that arise from the unfolding are only a minor contribution to
the total systematic error.

\subsection{Systematic Uncertainties}

Presently, the main systematic error of the energy spectrum reconstruction
comes from the
calibration ($7\prc$ in $E$). Also, in this preliminary study, there are still some
technical inaccuracies in the simulation, which
for instance lead to an incorrect reproduction of the signal threshold
function and consequently an inaccuracy of the likelihood function. These
technical issues contribute another $6\prc$ uncertainty in $E$.

Minor systematic errors come from the unfolding procedure,
and the statistical quality of the simulated datasets (each $2\prc$ in $E$).
In the CORSIKA shower simulation~\cite{corsika} two high energy interaction
models were tested up to now, namely SYBILL2.1~\cite{sibyll} and
QGSJet01.c~\cite{qgsjet}. The derived deviation in energy assignment
between the two models was
found to be less than $1\prc$, which is probably due to the low muon content
of the IceTop signal.

The sum of systematic errors is about $10-11\prc$, slightly depending on
energy. It is expected that most of the problems mentioned above
will be solved in the near future.

\begin{figure*} [t]
\centerline{\subfigure[Proton]{\includegraphics[width=0.95\columnwidth]{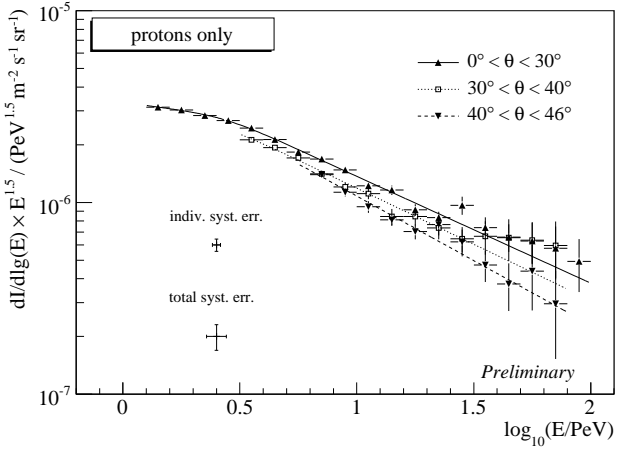}
\label{proton}}
\hfill
\subfigure[Iron]{\includegraphics[width=0.95\columnwidth]{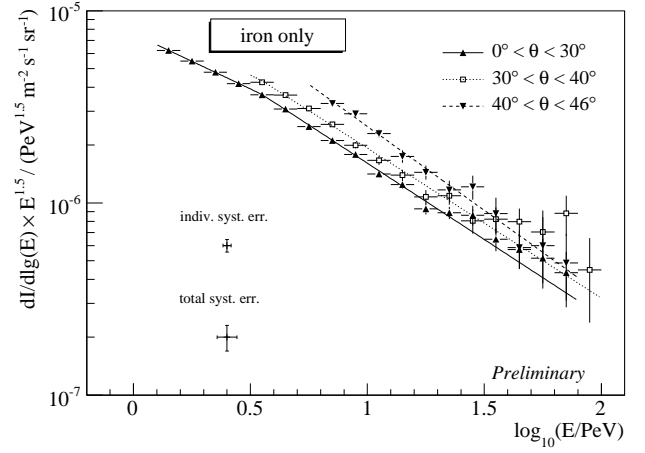}
\label{iron}}}
\centerline{\subfigure[Poly-Gonato]{\includegraphics[width=0.95\columnwidth]{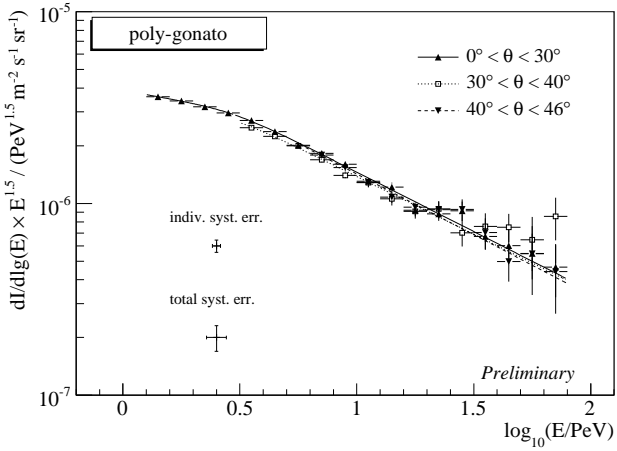}
\label{poly}}
\hfill
\subfigure[Two-Components]{\includegraphics[width=0.95\columnwidth]{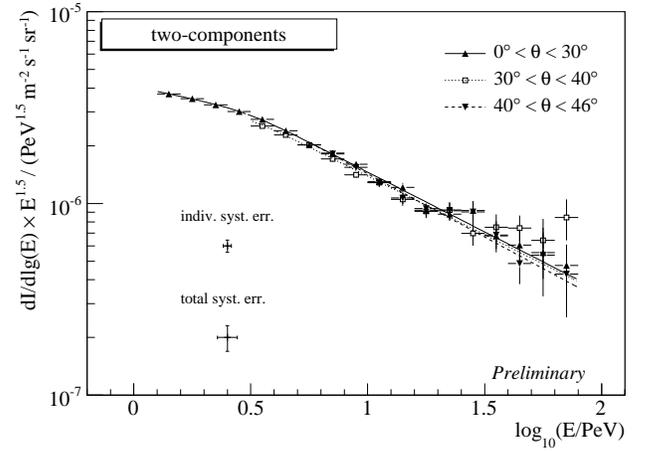}
\label{tc}}}
\caption{Preliminary, unfolded energy spectra for three zenith bands, assuming four
different composition assumptions. The shown points are those that are well
above threshold and that correspond to more than four
events~\cite{klepser_thesis}. The error bars of the single points represent statistical
errors. The total systematic error, and the error intrinsic to the inclination
bins, are displayed on the lower left (see text).}
\label{unfolding}
\end{figure*}

\subsection{Analysis of Three Inclination Ranges}

The air shower data recorded in August 2007 was subdivided into three zenith
bands that are roughly equidistant in $\sec\theta$, namely $\Omega_0 = [0\dg,30\dg]$,
$\Omega_1 = [30\dg,40\dg]$ and $\Omega_2 =[40\dg,46\dg]$. For each of these
bands, a proton and an iron response matrix were simulated. In addition,
two mixed composition response matrices were calculated.
One is a two-component mixture of protons and iron~\cite{two_components};
the iron
fraction increases from $34\prc$ at $1\eh{PeV}$ to $80\prc$ at
$100\eh{PeV}$. The other one is a 5-component implementation of the
poly-gonato model that turned out optimal
in~\cite{hoerandel_ontheknee}. Here, the elements above helium increase from
$40\prc$ to $98\prc$ in the same range.

In both cases, the mixed composition
matrices were
calculated as a superposition of the proton and iron responses. Using them for
unfolding means that only the relative composition goes into
the analysis, not the absolute flux scales of the models.

Figure \ref{unfolding} shows the energy spectra resulting from
the unfolding for the four response
matrices. The pure proton and iron assumptions lead to deviating spectra
with opposite ordering for protons and iron. Furthermore, the proton
spectra diverge towards higher energies, whereas the iron
spectra converge. This suggests that the response matrix needed for a
isotropic flux must be generated assuming a mixed composition with a mean mass
increasing with energy. In fact, the spectra obtained with the poly-gonato and
two-components models do agree much better.

\subsection{Results on Composition}

To quantify the observed discrepancy of the unfolded spectra, likelihood
values were
calculated that characterise the compatibility of the spectra. The most
sensitive method is to compare the values from the three zenith ranges with
their mean for each individual spectrum bin.
Since the absolute likelihoods rather characterise the statistical quality of
the dataset than the model itself, likelihood ratios were taken to validate
the models against
each other.
In this
comparison, care was taken to distinguish between systematic errors that apply
on all zenith bins equally (e.g. the muon calibration error) and errors that
do or may apply on the zenith bins independently.

The likelihood ratios with respect to the poly-gonato model were $4\ttt{-8}$
for pure proton and $2\ttt{-14}$ for pure iron composition, respectively. This
excludes both of the pure composition assumptions.
No preference could
clearly be identified between the two mixed composition models.

Although this finding is as yet not surprising, the power of it may increase
considerably as systematic and statistical errors will be reduced in the near
future. Furthermore, the benefit of this analysis is that it is 
complementary to the coincident measurement, since it exploits only the
development of the (mainly electromagnetic) showers and therefore is less
dependent on the production mechanisms for highly energetic muons.

\subsection{Results on the Energy Spectrum}

\begin{figure}
\centering
\includegraphics[width=0.92\columnwidth]{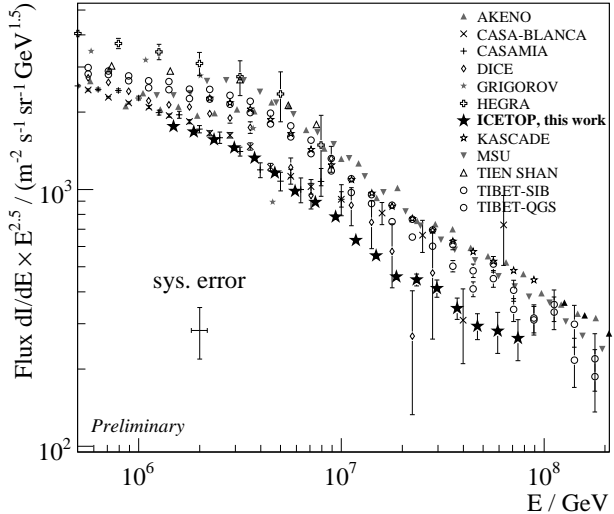}
\caption{Preliminary energy spectrum from $1-80\eh{PeV}$ measured by IceTop in
August 2007,
in comparison to results from other experiments. The error bars of the single
points represent statistical
errors. The total systematic error is displayed on the lower left.}
\label{spectrum}
\end{figure}

Figure \ref{spectrum} shows several energy spectra from other experiments,
along with
a preliminary spectrum from
IceTop, assuming the 5-component poly-gonato composition
model. The two-components model delivers almost the same result and is equally
qualified by the derived probabilities, so this choice by now is arbitrary. The following systematic errors are
given for the context of the poly-gonato
composition assumption, so they do not assess a possible deviation from that.

The spectrum can be fitted with a broken power law ($\chi^2/\mathrm{n.d.f.}=
9.5/13$). It determines the knee position at
$3.1\pm0.3\,(\mr{stat.})\pm0.3\,(\mr{sys.})\eh{PeV}$ and a power index change
from $\gamma_1=-2.71\pm0.07\,(\mr{stat.})$ to
$\gamma_2=-3.110\pm0.014\,(\mr{stat.})$. The preliminary estimate of the
systematic uncertainty of the power indices is $0.08$.

The absolute
flux, or energy assignment, is below that of most other spectra. Taking into
account the systematic error of our and the other measurements,
however, the deviation corresponds to no more than about
$1.5\,\sigma_{\mr{sys.}}$.

The low flux, or energy assignment, is a feature that is already found in the
energy distributions before the unfolding. Simulation improvements in the near
future will reduce the systematic error and probably clarify whether the
reason of this deviation is physical or not.

\section{IceTop-IceCube Coincident Analysis}

Detecting events with IceTop and IceCube in coincidence can be used to do a
composition analysis, but also to improve the event reconstruction. Both of
these efforts are still under development, but will make IceCube a
three-dimensional air shower detector in the near future.

\subsection{Reconstruction of Coincident Events}

Air showers near vertical, with the shower axis contained in both IceTop and
IceCube, can be observed in coincidence.
The signal in IceCube is caused by a muon bundle that usually has a spread of some
tens of meters, which is much less than the grid constant ($125\eh{m}$). This means that the single muon reconstruction algorithms used
in the neutrino analysis of IceCube can in principle be
applied to air shower data and lead to a good estimation of arrival direction.
Existing simulations indicate that a combined IceTop-IceCube reconstruction
may improve the overall shower direction resolution. 

A muon bundle reconstruction has to consider the ice properties and
longitudinal development of the muon number. It can lead to an estimation of size, i.e. muon
content, and the spread of the bundle and its light in time and space.
Adding the IceTop size and $\beta$ parameters, a coincident event is then
characterised by at least 4-5 parameters and has only two variables to be
determined, namely energy and mass. This allows for several reconstruction and analysis
approaches that at present are still
under development. 

\subsection{Analysis of Coincident Events}

A well-known quantity that is related to the primary mass of an cosmic ray air shower
is the ratio of electromagnetic to muonic particles ($e/\mu$). Heavy nuclei tend to
produce more muons and in addition develop faster, which mostly leads to a lower
$e/\mu$ on ground level.

The limiting issue in $e/\mu$ analyses is still the
understanding of the early high-energetic interactions that strongly
affect the muon production. It is therefore of great importance to
have experiments that detect air showers in orthogonal approaches.
Unlike many other experiments, IceTop has the ability to
complement its almost dominantly electromagnetic signal at the surface with a
measurement of the exclusive and highly energetic muon bundle in the deep IceCube detector. 

\begin{figure}
\centering
\includegraphics[width=0.90\columnwidth]{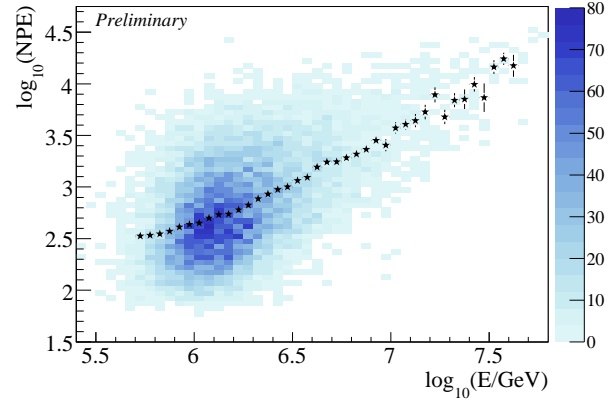}
\caption{
Coincident air shower events from experimental data. NPE is the sum of all recorded photons in IceCube. The
shaded histogram indicates the spread of the data, the stars are
the average values, displayed only for sufficiently populated bins.}
\label{coincident}
\end{figure}

Figure~\ref{coincident} shows experimental data of photon numbers in IceCube vs.\ reconstructed
energy in IceTop. 
As expected, the muon bundle size, related to the IceCube photon number,
clearly increases with energy.
Simulations show that the mean
signals of the two extreme cases of proton and iron showers are significantly
separated in this graph.
As in other experiments, the strong variations, intrinsic to the hadronic shower cascades,
require a statistical analysis of the data, probably involving unfolding and/or
sophisticated event classification techniques.

\section{Surface Muon Counting}

\begin{figure}
\centering
\includegraphics[width=0.92\columnwidth]{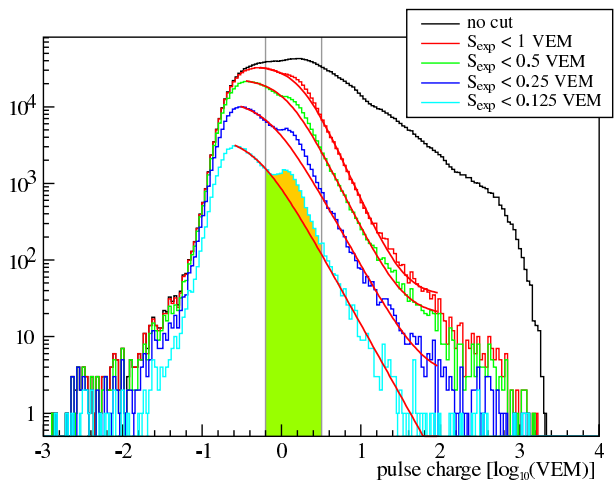}
\caption{Pulse charge distributions in experimental data
for different maximal expectation value conditions.
At lower expectation values, i.e. high distances from the shower axis, a peak
becomes visible at about $1\eh{VEM}$ that corresponds to single muons hitting
a tank (filled area) \cite{diplomarbeit_adam}.}
\label{muon_counting}
\end{figure}

Although IceTop records light curves in high precision in the tanks, muon signals are
difficult to identify due to the quantitative dominance of
electromagnetic particles. However, at large distances from the shower core,
where the overall charge expectation is well below $1\eh{VEM}$, single muons
can produce bright signals that can be used to estimate their abundance in a
statistical way. In 2007, the
array was already big enough to identify such muons
(Fig.\,\ref{muon_counting}). 

This can be a twofold benefit: First, the number of muons, or an estimator
for it, can be used to do a composition analysis and scrutinise interaction
models. A first work that is still in progress is comparing muon peak heights in
data with those in simulations. It already reproduces the
effect in general, and shows  quite a large difference between proton and iron
simulations, which suggests a good sensitivity for a composition analysis and
model testing.

Secondly, the muon peak may be used to do an online
monitoring of the calibration data, complementing the muon calibration runs
that are currently done on a regular basis in between the data runs.

Also under study is the identification of highly
energetic muons with high transverse momenta in deep IceCube. These may be seen
far from the main muon bundle and deliver information about high-$p_t$
particles; the interactions that produce these particles may be understood in
a perturbative QCD context \cite{icetop_highptmuons}.

\section{Conclusion}

The IceTop air shower array at the South Pole is half completed and
continuously taking physics data. Shower reconstruction algorithms have been
developed and tested. They lead to competetive
resolutions in shower direction, core position and primary energy.

A first study of the energy spectrum, using IceTop as a standalone detector
and the data from one month in 2007,
yielded two results: First, a sensitivity on cosmic ray composition
was found by comparing energy spectra from different inclinations. A first
study, using pure proton, pure iron and two mixed modellings of cosmic
rays, showed a clear
preference for the two mixed composition models. Secondly, an energy
spectrum between $1-80\eh{PeV}$ was extracted that shows all expected features, and, within
uncertainties, agrees relatively well with other measurements.

The reconstruction and analysis of IceTop/IceCube
coincident events is still under development. IceCube offers various
possibilities to interpret the three-dimensional shower images, making use
of time and signal height information on the surface and deep in the ice.

A new analysis is being developed that aims at the identification of single
muon signals in IceTop, at large distances from the shower axis.
This will lead to another, yet complementary composition analysis method and
may be usable for testing air shower models.

\bibliographystyle{IEEEtran.bst}
\bibliography{icetop}
%
%
%

\end{document}